# China's First Workforce Skill Taxonomy

Weipan Xu[1†], Xiaozhen Qin[1†], Xun Li[1*], Haohui "Caron" Chen[2,3*], Morgan Frank[4], Alex Rutherford[5,6], Andrew Reeson[2], Iyad Rahwan[5,6*]

**Affiliations**

[1]Department of Urban and Regional Planning, School of Geography and Planning, Sun Yat-sen University, Guangzhou, China

[2]Data61, Commonwealth Scientific and Industrial Research Organisation (CSIRO), Australia

[3]Faculty of Information Technology, Monash University, Australia

[4]Institute for Data, Systems, and Society, Massachusetts Institute of Technology, USA

[5]Center for Humans & Machines, Max-Planck Institute for Human Development, Lentzealle 94, Berlin 14195, Germany

[6]Media Laboratory, Massachusetts Institute of Technology, USA

*Correspondence to: Xun Li (lixun@mail.sysu.edu.cn), Haohui "Caron" Chen (caronhaohui.chen@data61.csiro.au), and Iyad Rahwan (rahwan@mpib-berlin.mpg.de).

† These authors contributed equally to this work.

## Abstract

China is the world's second largest economy. After four decades of economic miracles, China's economy is transitioning into an advanced, knowledge-based economy. Yet, we still lack a detailed understanding of the skills that underly the Chinese labor force, and the development and spatial distribution of these skills. For example, the US standardized skill taxonomy O*NET played an important role in understanding the dynamics of manufacturing and knowledge-based work, as well as potential risks from automation and outsourcing. Here, we use Machine Learning techniques to bridge this gap, creating China's first workforce skill taxonomy, and map it to O*NET. This enables us to reveal workforce skill polarization into social-cognitive skills and sensory-physical skills, and to explore the China's regional inequality in light of workforce skills, and compare it to traditional metrics such as education. We build an online tool for the public and policy makers to explore the skill taxonomy: skills.sysu.edu.cn. We will also make the taxonomy dataset publicly available for other researchers upon publication.

# 1. Introduction

Workers rely on their skills to earn a living. Fluctuations in the demand for particular skills translate to changes in wages and employment opportunities for individual workers. Technology has long driven such changes, which can result in significant economic and social upheavals[1]. Technology acts as a substitute for some skills, for example calculation, but complements others, for example accountants who use the outputs of those calculations; in general, routine work has proven most susceptible to substitution, while non-routine tasks have been complemented by technology[2]. For those with complementary skills, technology increases their productivity, and hence the wages their skills are likely to command[3]; for those without such skills the outlook is not so good.

Recent decades have seen particularly rapid technological change, and there is considerable evidence that technology, and related changes in global trade, are impacting labor markets. For example, in the US, the 'college premium', the wage advantage enjoyed by more educated (skilled) workers, has increased over the last 40 years while at the same time low skilled workers have seen their real earnings decline[4]. Over a similar time period many European countries also show evidence of job polarization, with declines in the middle third of the job distribution[5]. Unfortunately, it is not a simple matter for workers to move from declining to growing occupations. A recent analysis of US skills data shows a marked polarization between two distinct clusters of skills, one (social-cognitive) associated with high-income jobs and the other (sensory-physical) with low-income jobs[6].

With 758 million workers, China represents by far the world's largest labor market. Despite relatively low wages by international standards, automation is occurring increasingly rapidly in China[7] which has the potential to cause significant disruption. Understanding the distribution of skills across the workforce is important to help manage these impacts on individual workers and their communities. However, China lacks a skill taxonomy, which limits the assessment of the distribution of skills across occupations and regions. China's National Occupation Classification Code (NOCC) includes occupation titles and corresponding task descriptions. By contrast, the US Department of Labor's Occupational Information Network (O*NET) describes the importance of 161 skills, knowledge and abilities for each of its 672 recognized occupations.

The aim of this paper is to build China's first workforce skill taxonomy and apply to it test for evidence of the skills polarization observed elsewhere. We build a Naïve Bayes Model to infer the relationships between O*NET's occupational tasks and skills, and use the model to predict skills needed across Chinese occupation based on their job descriptions. The resulting skill taxonomy can provide a new perspective for studying the growing regional inequality in China, if we view cities as "abstract bundles of workplace tasks and skills" as well[2]. Chinese city's industrial structure varies extensively. Major cities in China have transformed from manufacturing hubs to service centers, while large numbers of medium and small cities are still specialized in mining, manufacturing and farming. The heterogeneous industrial structure, if regarded as "tasks", infers distinct city skill profiles, in which large cities might have significant social-cognitive skill stock, while medium and small cities have sensory-physical skill stock. The thorough understanding of city skill profile helps policy makers better understand how skill profile determine the city economic performance and accordingly reskill workers for adapting to emerging changes. Therefore, we used the skill taxonomy to compute the city skill profile and explain the regional inequality of economic growth and identify the polarization of Chinese workforce skills. The China's first workforce skill taxonomy and the findings based on that could open a new research agenda for China's labor market research.

# 2. Results

## 2.1 China's polarized labor market

Labor market polarizations have been observed in the U.S. and many European countries since the 1980s[2,6]. Chinese scholars also started to pay attention to China's labor market polarization problem in recent years. Lacking the occupation-specific data, scholars can only draw conclusions upon the macro data like employment data and manufacturing sector data, or the microdata like workforce survey, leading to two opposite findings, one[9] observed polarization while the other[10] couldn't ascertain. Based on the skill taxonomy, we revisit the question of whether or not the polarization exists in China, which is of significance for policymakers.

Fig. 1a addresses the skill distributions over six major occupational groups, demonstrating extensive heterogeneities, in which white collars like Department and Enterprise Heads and Technicians and Professionals heavily depend on soft skills such as *social*, *mental process* and *complex problem-solving,* while blue collars like Manufacturing Workers depend on hard skills such as *Psychomotor Abilities*, *Work Output,* and *Technical Skills*. Moreover, the numbers of skills also vary significantly across occupations, with some occupations requiring more than 80 skills while some require less than 40 skills (Fig. 1b). The Technicians and Professionals require the highest average numbers of skills at 71 while Manufacturing Workers and Agricultural Workers both require the least at 55 and 58 respectively (Fig. 1c). Moreover, while the latter two major groups account for the most significant shares of employment at 23% and 47% respectively, and their skill contents are primarily composed of skills that are susceptible to automation, the technological unemployment risks are elevated. Besides, the majority of Manufacturing Workers and Agricultural Workers reside in small and medium cities, in which economies rely on one or a few industries and are relatively weaker than megacities. It is more challenging for small and medium cities to implement policies to alleviate automation impacts such as career migration and reskilling.

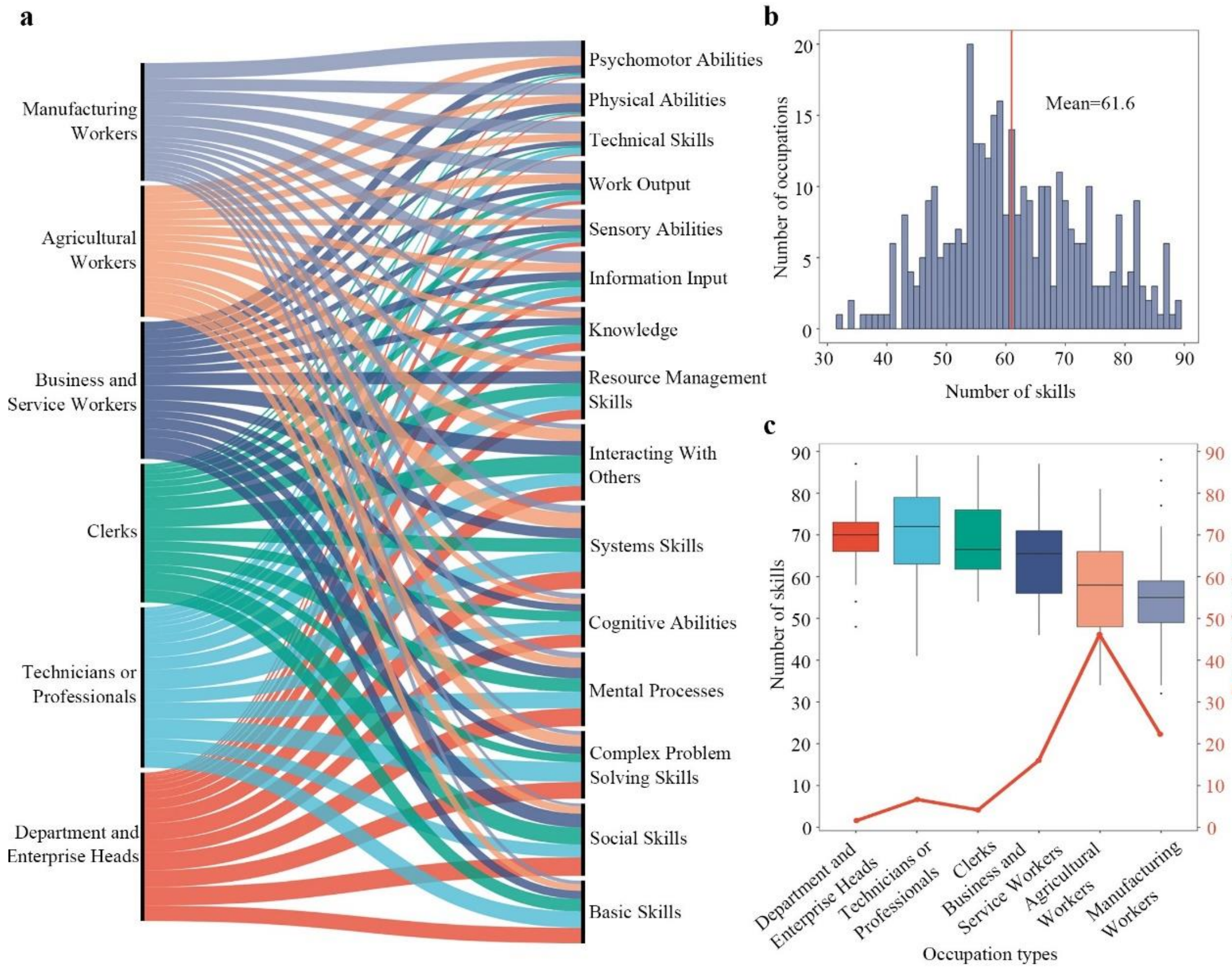

**Fig. 1 | Skill distributions over Chinese occupations. a**) The Alluvial Diagram indicates the degree to which the occupational groups depend on different skill sets. We aggregate the skill taxonomy into the 6 major occupational groups, and 15 skill sets defined by the O*NET. The thickness of the line is proportional to the aggregated values of skill importance. **b**) The frequency distribution shows skill distributions across all 353 Chinese occupations. The vertical red line addresses the average numbers of skills that are required by Chinese workers. **c**) The box-plot shows the numbers of skills that are important to the six major occupation groups; the red line shows the corresponding employment shares.

China has adopted most industrial robots in the world since 2016[7], which are believed to replace large number of low-skilled manufacturing workers. A worker might benefit from reskilling to gain more career mobility opportunity. For instance, according to the World Economic Forum, with appropriate reskilling, there exists a career pathway for assembly line workers transiting to construction laborers, and First-Line Supervisors of Construction Trades and Extraction Workers[11]. Therefore, we built a skill space (Fig. 2a) that addresses relationships between skills to understand the career mobility opportunities for all occupations. The skill-pair proximity (the edge of the skill space) is the minimum probability of the corresponding skill-pair $\theta\left(s, s'\right)$ co-occurring in the same occupations as depicted in Eq. [1]. We denote whether or not skill $s$ is important in occupation $o$ as

$$\theta\left(s, s'\right) = \frac{\sum_{o \in O} e(o,s) \cdot e(o,s')}{\max\left(\sum_{o \in O} e(o,s), \sum_{j \in J} e(o,s')\right)} \quad [1]$$

We identified two skill clusters using *fast unfolding* in the skill space[12], revealing the skill polarization, in which skills like *Knowledge, Social* and *Cognitive* constitute a socio-cognitive cluster and skills like *Physical, Sensory* and *Work Out* constitute another sensory-physical cluster. The skill polarization in China aligns with the findings in the U.S.[6], even though the skills linking two clusters, that is, the bridging nodes, is significantly different. The transition corridor between the two clusters comprises of skills like *Mathematics, Judging the Qualities of Things, Services, or People,* and *Estimating the Quantifiable Characteristics of Products, Events, or Information* (Fig. 2a). Workers can master the last two skills by accumulating work experiences, but *Mathematics* can only be derived through systematic learnings, e.g., attending higher education. However, China's 4-year degree or higher education attainment rate at age 25-34 is only 14% in 2018, far behind OECD countries and other developing countries[13], which means the reskilling and career mobility is challenging for most blue-collar workers. Besides, *Mathematics* is a vital complementary skill to *Science, Physics* and *Programming* (highlighted in Fig. 2a), and all four skills are integral parts of STEM (Science, Technology, Engineering and Mathematics) defining a nation's competitiveness[14]. In China's labor market, they also work as bridging nodes or obstacles of career mobility.

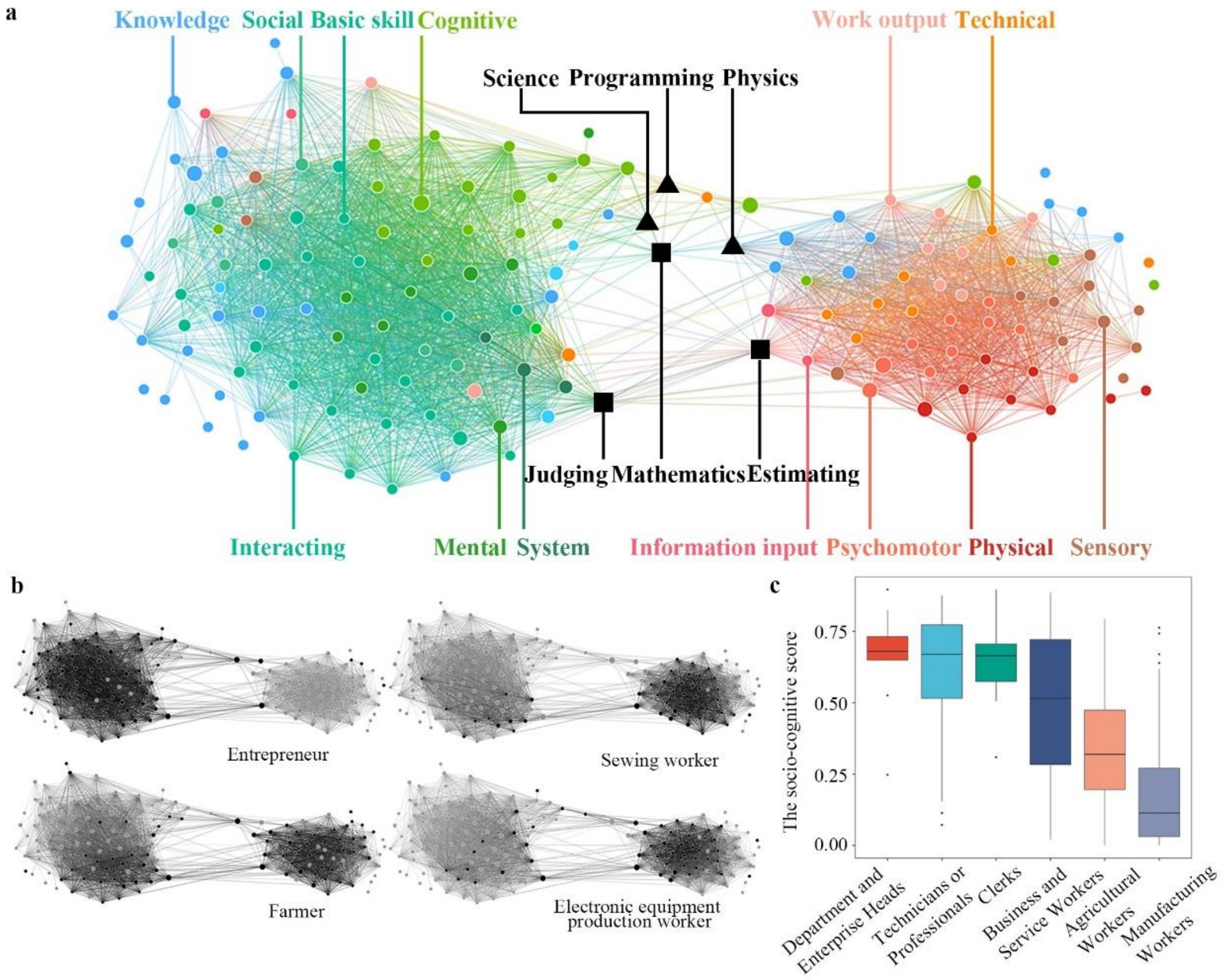

**Fig. 2 | China's workforce skill space. a**) A skill space addresses the relationships between skill pairs. The edges represent skill-pair proximities that are the minimum probability of the corresponding pair co-occurring in the same occupation. Node colors represent the skill categories defined by O*NET and node sizes reflect the betweenness centrality. We name the cluster on the left "socio-cognitive" skill cluster and the right "sensory-physical" skill cluster, given the types of skills that constitute them. Three skills *Judging the Quality of Things, Mathematics*, and *Estimating the Quantifiable Characteristics of Product, Event or Information*, are highlighted in black squares, as they have the greatest betweenness centrality in the network indicating their transition positions between the two clusters. *Mathematics* is a complementary skill to *Science, Physics* and *Programming* which are highlighted in the black triangles. **b**) Four occupations' positions in the skill space. The black nodes represent the skills that are important to the corresponding occupations. **c**) The socio-cognitive scores of 6 major occupation groups. Department and Enterprise Heads and Technicians and Professionals have greater socio-cognitive scores than the other occupation groups.

While highlighting some occupations' positions on the skill space (Fig. 2b), we found distinct skill sets between the white- and blue-collar workers. For instance, the *entrepreneur* has 66 out of a total of 72 skills belonging to the socio-cognitive cluster, while the *farmer* has 40 out of a total of 71 skills belonging to the sensory-physical cluster. Reskilling the latter is of particular significance, as it accounted for 41% of the total employment in 2010. Moreover, China's rapid urbanization, which would exceed 70% in 2030[15], would further suppress the demands of this workforce. Fortunately, *farmers* are on top of a few socio-cognitive skills like *Systems Evaluation* and *Management of Personnel Resources*, which would make their career mobility feasible. Over the past decades, industries such as electronic equipment production and textiles have made major contributions to China's industrialization. The most representative jobs in these two industries are Electronic Equipment Production Workers and Sewing Workers, accounting for 0.3% and 1.7% of total employment respectively. Both have skills concentrated on the sensory-physical clusters, so reskilling them is challenging yet of significant importance. Computing career pathways for specific occupations is out of the scope of this study, but the skill space provides a good starting point.

From the examples above, one might assume blue-collar workers could be reliant on the sensory-physical skills and white-collar workers on socio-cognitive skills. Therefore, we compute the occupation-specific socio-cognitive score based on the proportion of the socio-cognitive skills to a job's overall skills (see Eq. [3] for the computation and Data S3 for socio-cognitive scores for each occupation) and confirm such an assumption (Fig. 2c). In the U.S, jobs that have higher socio-cognitive scores tend to have higher wages[6]. Even though China's wage statistics is only available at the major group level, we still observe similar findings. The socio-cognitive jobs like Managers and Professionals earning respectively CNY$131,929 and CNY$83,148 per year are significantly more than sensory-physical jobs like Business and Service Workers and Manufacturing Workers earning respectively CNY$49,502 and CNY$50703 per year. The Clerks have relatively high socio-cognitive scores, but they just earn CNY$58,211 per year modestly more than the last two occupation groups (the wage statistics uses different occupation coding than the NOCC, but they are still comparable. See Table S2 for the detailed wage statistics and explanation). The polarized structure of the skill space provides us with a new perspective for studying China's wage inequality problem, in which unequal access to education resources is only one part of the story[16], and the distinct skill sets between the high- and low-paid workers have always been neglected. The skill space also informs two reskilling pathways to bridge the wage gap. However, learning *Mathematics* and all other STEM skills might not be feasible for the overall under-educated workforce unless a more inclusive education system, e.g., vocational training, is implemented.

## 2.2 City skill profile and its effects on economic growth

If we view a job as a bundle of tasks and a task requiring a series of skills[4], a city, as a container of jobs, can be viewed as bundles of tasks and skills. In this part, we build the Chinese city skill profile using the skill taxonomy and study the regional inequality problem from the perspective of workforce skills.

A city skill profile is defined by two parts; one is the numbers of skills effectively used by the city c, i.e., $Skill_c$, (see Eq.[8]), and the other is the proportion of socio-cognitive jobs that the city has, i.e., $Cognitive_c$ (see Eq. [10]).We determine whether or not a particular skill is effectively used by a city by its importance to the corresponding city over the importance of that skill to all the cities. The numbers of skills used effectively by cities range from 55 to 103 out of a total of 161 skills. The range between 70-80 skills accounts for more than half of the cities (Fig. S3). and $Skill_c$ might represent a city's skill diversity and we found a modest correlation between it and the GDP per capita (Model 3 in Table 1).

In terms of skill content, some cities like Guangzhou, China's third largest city, dominate in socio-cognitive skills while the others like Putian, famous for its shoemaking industry, mainly rely on sensory-physical skills (Fig. 3a). The city skill profile might address China's underlining distinct industrial structure, where service centers rely on socio-cognitive skills while manufacturing, mining and farming hubs rely on sensory-physical skills. If a city has a large number of jobs that depend on socio-cognitive skills, which are more resilient to digital disruptions, one might wonder if it also has stronger economic performance. Therefore, we calculate the city's cognitive scores by the proportion of jobs that mainly rely on socio-cognitive skills (Eq. [9] and [10] in appendix) and use it to explain the economic performance. Beijing has the highest score of 0.52 while Nanyang has the lowest score of 0.09 (see Fig.S4 and Data. S2 for cognitive scores of all cities). Moreover, the cognitive score is consistent with the education levels: a city of a relatively high cognitive score also has a relatively large number of well-educated workers (Fig.S4).

Given education levels were used as a main explanatory variable for explaining economic performance[17,18], we use it as the benchmark to explore whether or not skills have stronger explanatory power (Table 1). Model 1 and model 2 address the effects of the education levels and the cognitive scores on urban economic performance (measured as per capita GDP) respectively. The results show that both could benefit economic growth, but the cognitive score ($R^2$=0.5145) has better explanatory power than the education levels ($R^2$=0.3094). When both variables are considered simultaneously in Model 4, only the cognitive skill level is significant. Therefore, city skill profile has stronger explanatory power in explaining economic performance than the conventional explanatory variable education factor. Additional regression analyses on per capita wage shows similar findings (Table. S3). Interestingly, we found that "sub-provincial or above" cities, which enjoy higher administrative powers than their peers, tend to have higher cognitive scores (Fig. 3b). The "sub-provincial or above" cities, largely comprised of the provincial capitals, are a legacy of the planned economy, in which a city of higher administrative power tends to own more premium resources such as centrally-funded universities and state-owned corporation. Those cities normally serve as service hubs within the region for the surrounding areas which are dedicated to manufacturing, mining and farming. In this regard, we might be able to explain China's regional inequality from a new geography and skill combined perspective, in which there exists a core-periphery layout where provincial capital cities provide socio-cognitive skills and periphery cities provide sensory-physical skills.

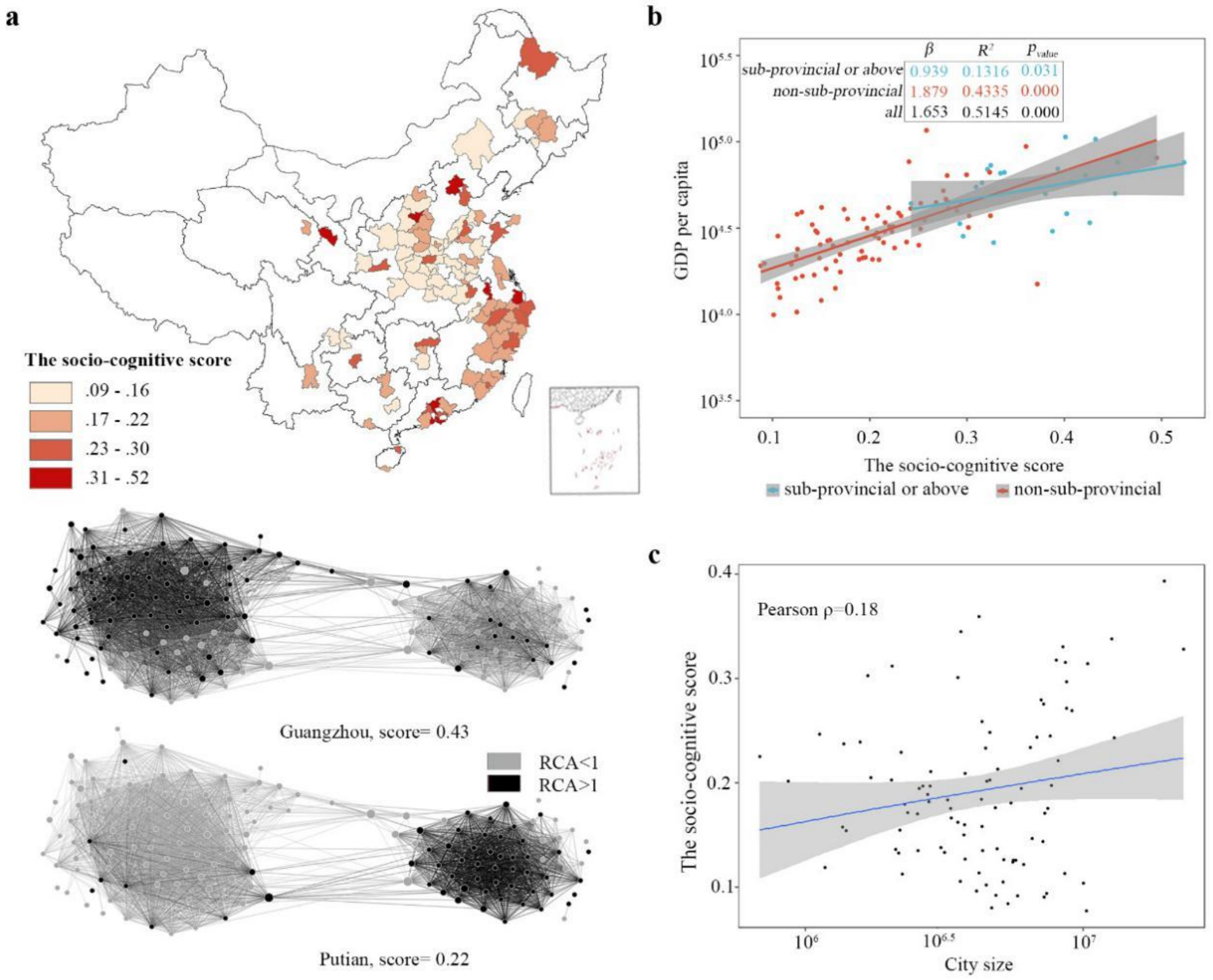

**Fig. 3 | City skill profile. a**) The skills that are effectively used by a city are highlighted in the skill space. Two cities representing high- and low-level cognitive scores were selected as examples. **b**) Strong correlation is found between the GDP per capita and the city cognitive scores. Cities of higher administrative power, so called "sub-provincial or above" cities, highlighted in blue, tend to have greater cognitive scores and GDP per capita than their peers. **c**) The scatter diagram shows a weak relationship between city's cognitive scores and its population size.

**Table 1. Estimating effects of education and skill profile on urban economic.**

| | (1) | (2) | (3) | (4) |
|---|---|---|---|---|
| | GDP per capita | GDP per capita | GDP per capita | GDP per capita |
| **University degree** | 3.479*** | | | -0.0465 |
| | (5.68) | | | (-0.06) |
| **Socio-cognitive score** | | 1.653*** | | 1.667*** |
| | | (9.59) | | (5.06) |
| **Skill number** | | | 0.014*** | |
| | | | (6.85) | |
| **_cons** | 4.373*** | 4.125*** | 3.464*** | 4.124*** |
| | (136.91) | (99.10) | (22.48) | (79.49) |
| **N** | 102 | 102 | 102 | 102 |
| **R2** | 0.309 | 0.515 | 0.334 | 0.515 |

*t* statistics in parentheses; GDP per capita is measured in 2010; Skill number is the number of skill effectively used in each city. $^{*}$ $p < 0.05$, $^{**}$ $p < 0.01$, $^{***}$ $p < 0.001$

## 2.3 Skill's effects on labor immigration

China's regional inequality also stems from the distinct attractions of workers amongst cities, reflected by labor immigration. Population growth has been an important indicator for the local governments to estimate the economic performance from the mid 1990s[19] when China started to enter its manufacturing export prime. Job opportunity is the primary motivation for immigration. Assuming job opportunities are proportional to the employment size, radiation model is applied to accurately predict the migration patterns amongst U.S. counties[20]. However, in China, job opportunity is not necessarily proportional to the employment size due to the extensive heterogeneity in local labor market structure. In contrast to the U.S.[6], employment size has nearly no correlation with one's socio-cognitive score in China (see Fig. 4c). For example, despite Shenzhen and Nanyang own labor markets of similar size at 10 million workers, their socio-cognitive scores differ significantly, at 0.31 and 0.07, respectively. Therefore, assuming a worker would get similar level of job opportunity at cities of similar employment size would not be an appropriate prior of apply radiation model in China. Instead, we assumed skilled population size is the major human capital of a city, given cities with more skilled workers tend to attract more immigrants[21,22]. Therefore, we adapt the radiation model assuming job opportunities are proportional to a city's number of skilled workers to test whether or not migration patterns between cities are better explained. To be more specific, we input the number of college-educated workers, skilled workers and overall employment respectively to the radiation model (Eq. 2), deriving the predictive number of labor immigrants between each two cities. We denote the ratio of immigrants from city $i$ into city $j$ to all immigrants from city $i$ as

$$T_{ij} = \frac{m_i n_j}{(m_i+n_j)(m_i+n_j+s_{ij})} \qquad [2]$$

where $m_i, n_j$ is the population of city $i$, $j$ respectively. $s_{ij}$ is the total population in circle of radius $s_{ij}$, distance from city $i$ to $j$, centered at city $i$. Population data is represented by the number of skilled workers, all workers one by one.

We collected a daily city-to-city migration data (at July 24th, 2019) from Baidu Map® to verify the accuracy of the prediction. As the migration data only shows the top 10 destination cities for each city, we use the Normalized Discounted Cumulative Gain (NDCG) to measure to what extent the radiation model can predict the migration patterns. As showed in Fig. 4a, the average NDCG values of radiation model based on the skilled worker, the educated worker and the overall employment size are 0.65, 0.67 and 0.61 respectively. It turns out that the model based on skilled worker size is not significantly different from the one based on college-educated worker ($p_{t-test} = 0.19$). However, it is significantly better than the model on the overall employment size ($p_{t-test} = 0.02$). Taking Tianjin as an example (Fig. 4b, c, and d), both radiation models (using employment size and skilled workers) can successfully predict 7 out of the top 10 destinations. However, when it comes to the rank accuracy, skilled worker model performs better than the baseline model. Specifically, the skilled worker model predicts Beijing as the very top destination, aligning perfectly with the migration data, while the baseline model predicts Langfang, which only ranks 4th in reality. Despite Tianjin being closer to Langfang than Beijing, more workers will migrate to Beijing due to the latter's larger skilled worker size. Therefore, skilled population could be a better indicator of job opportunity or city attractiveness to predict the labor migration.

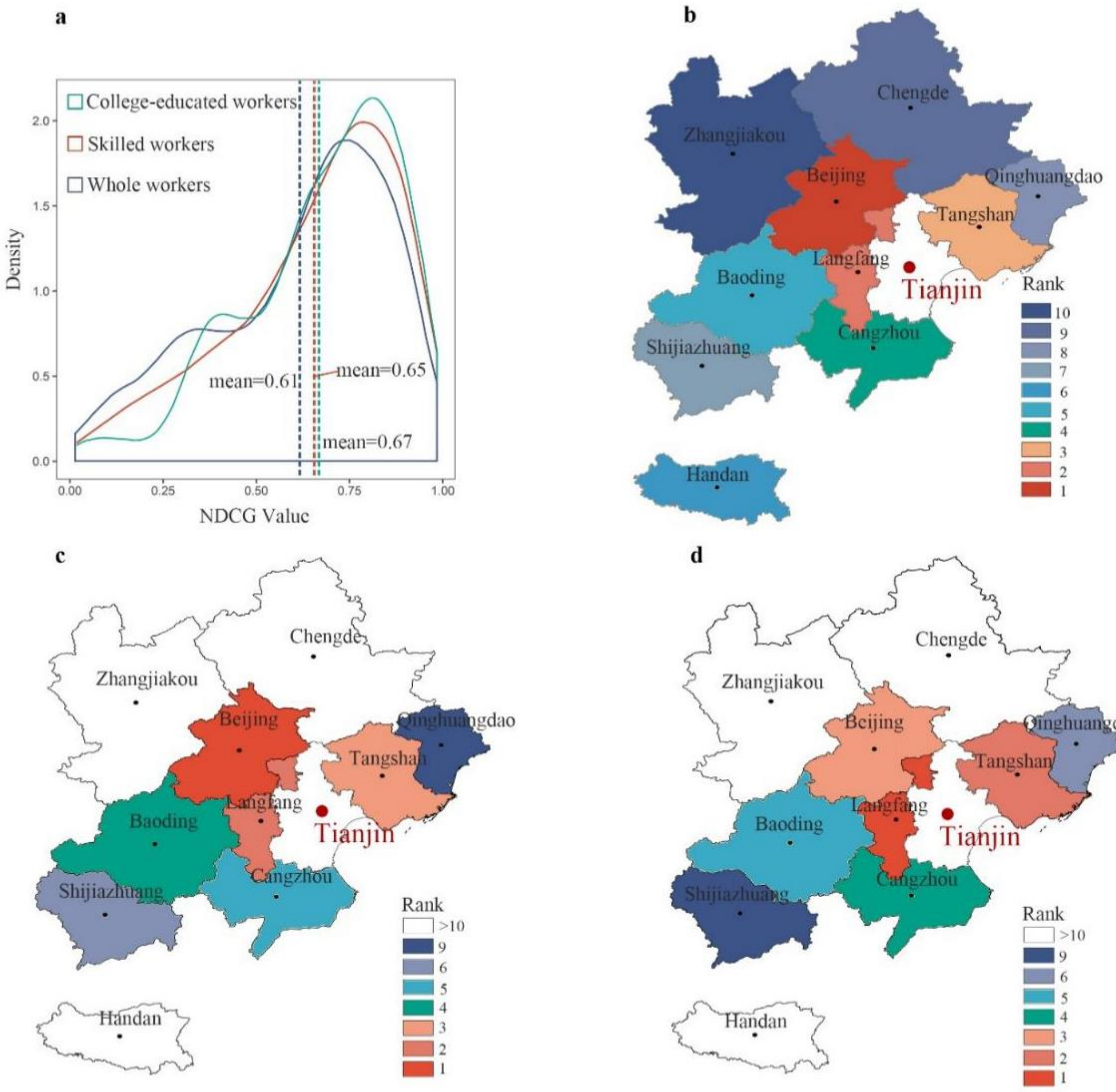

**Fig. 4 | Radiation models on predicting city-to-city migration patterns. a)** The distribution of NDCG values of the radiation models based on college-educated worker, skilled workers and overall employment size respectively. **b)** The top 10 destinations from Tianjin city based on the daily migration data from Baidu Map®. **c)** The predicted destinations from Tianjin city based on the baseline radiation model using employment size. **d)** The predicted destinations from Tianjin city based on the radiation model using skilled worker size.

# 3. Conclusion and Discussion

China's skill taxonomy could open a new agenda for researching both labor force and regional inequality issues in China. It is striking that our skills taxonomy better predicted city level GDP than even education, suggesting it is a useful research and policy tool. Applying the taxonomy across cities reveals considerable variation in skills profiles. Consistent with Alabdulkareem et al. for the U.S.[6], we find marked skills polarization across at both the occupational and geographical level in China. However, while in the USA it tended to be the smaller cities which had the lowest level of socio-cognitive skills, and hence incomes, the legacy of a centrally planned economy means that in China even relatively large cities can have low levels of these skills.

Over recent decades both employment and wage have tended to be higher for jobs involving socio-cognitive skills. Given the highly uneven distribution of these skills among Chinese cities, this suggests geographic inequality is increasing, which is illustrated by divergent population grow rates among cities (see Fig. S5). Those cities which are heavily dependent on socio-cognitive skills are likely to see higher wage per capita (see Table S4). The stark polarization of skills across occupations indicates that workers may struggle to transition from occupations in declining demand to those for which demand is growing. And even if they could transition, the lack of occupational diversity in many Chinese cities reduces the opportunities to do so.

Cities with a diversified skills base will be more resilient in the face of technological and other social and economic changes. Globally the decline of routine jobs such as in manufacturing has coincided with the growth of the service sector. However, service sector opportunities will be limited in cities with few highly skilled workers to drive demand. This suggests a key priority for policy will be to find ways to help both workers and cities find ways to transition from being dependent on the sensory-physical cluster of skills. A better understanding of occupations and skills in the Chinese context is an important step on this pathway.

**This PDF file includes:**

# 4. Material and Method

## 4.1 O*NET discussions

The O*NET database is updated yearly, providing information about the relationship among occupations, skills, and tasks. Besides, it provides information about the importance of *abilities*, *interest*, *knowledge*, *skills*, *work activities*, *work context* and *work values* of each occupation. The importance ranges on a scale from 1 (not important at all) to 5 (extremely important). Amongst, we use *abilities*, *knowledge*, *skills*, *work activities* to characterize occupational skills (161 in total). According to O*NET, 161 skills are categorized into 15 groups: *cognitive abilities, complex problem-solving skills,* *information input, interacting with others, knowledge, mental processes, physical abilities, psychomotor abilities, resource management skills, sensory abilities, social skills, systems skills, technical skills, work output,* and *basic skills.*

We denoted the importance of skill $s \in S$ to occupation $o \in O$ using $onet(o, s) \in [0, 1]$, where $onet(o, s) = 1$ indicates that $s$ is essential to $o$. However, some skills are ubiquitous amongst occupations, such as Identifying Objects. Therefore, adopting the work of Alabdulkareem et al[6]., we determine whether or not a specific skill is 'effectively used' by an occupation by calculating the corresponding revealed comparative advantage (*RCA*) as depicted in Eq. [3].

$$\mathrm{RCA(o,s)} = \frac{onet(o,s)/\sum_{s' \in S} onet(o,s')}{\sum_{o' \in O} onet(o',s)/\sum_{o' \in O, s' \in S} onet(o',s')} \quad [3]$$

We denote effective use of skill $s$ in $o$ using $e(o, s) = 1$ if $RCA\,(o, s) > 1$, and $e(o, s) = 0$ otherwise.

## 4.2 China's National Occupation Classification Code

NOCC provides us the main material to understand the occupation work content in China. Referring to the basic principles and the structure of International Occupation Standard Classification (IOSC) enacted by the International Labor Organization (ILO), China compiled two versions of NOCC by far in 1999 and 2015.

Compared with O*NET database, NOCC only define occupational titles and job descriptions lacking the quantitively measured skills and abilities. The classification has a hierarchy of four levels: major class, middle class, minor class, and unit class. The 2015 version has 8 major classes, 75 middle classes, 434 minor classes, and 1481 unit classes. The description of each occupation is the general statement of the definition, main work content and work process. Although lacking the information about workforce skill, we can extract task information from the job description. In order to be consistent with the classification of employment data in the 2010 census, we selected the 434 minor classes as the research material.

## 4.3 From Text to Skill (MIC)

A job can be conceptualized as a bundle of tasks[2,8], and a task requires a series of skills. Therefore, the relationships between skills and tasks can be inferred based on their co-location probability on the same job. We extract $K = 1273$ tasks from O*NET job descriptions in the form of word tokens (Fig.2a) and use $I = 161$ skills from O*NET to build a task-skill relationship matrix $A = (a_{k,i}) \in \mathbb{R}^{K \times I}$, where $a_{k,i}$ indicates the *mutual information* of the task-skill pair based on their distribution across O*NET's 696 occupations. The *mutual information* is a measure of stochastic interdependence between two variables[23]. The heat map (Fig.S1a) addressing the task-skill relationships demonstrates some tasks are highly related to some specific skills. For example, a skill called *Medicine and Dentistry*, is more likely to co-occur with tasks like *patient, treat,* *care, disease* and *medical* (see Fig.S2b for more examples). Therefore, we assume whether or not a certain occupation $o$ requires a certain skill $s$ can be inferred by the underlying tasks $t_{i,o} \in K$. Given that NOCC addresses $t_i$ of each Chinese occupation, we can infer the occupational skills, that is, $P(s^{O*NET}|o^{Chinese})$. To address the relationship between skills $s^{O*NET}$ and occupations $o^{Chinese}$, we decompose the conditional probability as Eq. [4] based on Bayes' theorem.

$$P(s^{O*NET}|o^{Chinese}) = \frac{P(s^{O*NET})P(o^{Chinese}|s^{O*NET})}{P(o^{Chinese})} \quad [4]$$

As occupations are a bundle of tasks, we derive the conditional probability $P(o^{Chinese}|s^{O*NET})$ as Eq. [5], where each $o^{Chinese}$ is regarded as a set of tasks $t_i \in \{0, 1\}$.

$$P(o^{Chinese}|s^{O*NET}) = \prod_{i \in K} P(t_i|s^{O*NET}) \quad [5]$$

A tripartite network (Fig.S2b) is built to link O*NET 1273 tasks, 161 skills and 696 occupations, so the probability of whether or not a certain task $t_i$ would occur given a set of skills $s^{O*NET}$ can be derived (Fig.S2c). Because a task $t_i$ is transversal across the U.S and Chinese workforce, the task-skill relationships $P(t_i|s^{O*NET})$ can be applied in the Chinese context, thus solving Eq. [5] and then Eq. [4] as shown in Fig.S2d. The resulting Chinese skill taxonomy (called skill taxonomy hereafter) addresses occupational skill distributions in binary values. For example, skills like *Finger Dexterity, Equipment Maintenance* and *Near Vision* are important to *sewing workers* but not so important to *entrepreneurs*. To the best of our knowledge, it's the first time that Chinese occupations can be quantitively measured in terms of skill importance. Details about skill distributions of all Chinese occupations can be found in Data. S1 in the appendix.

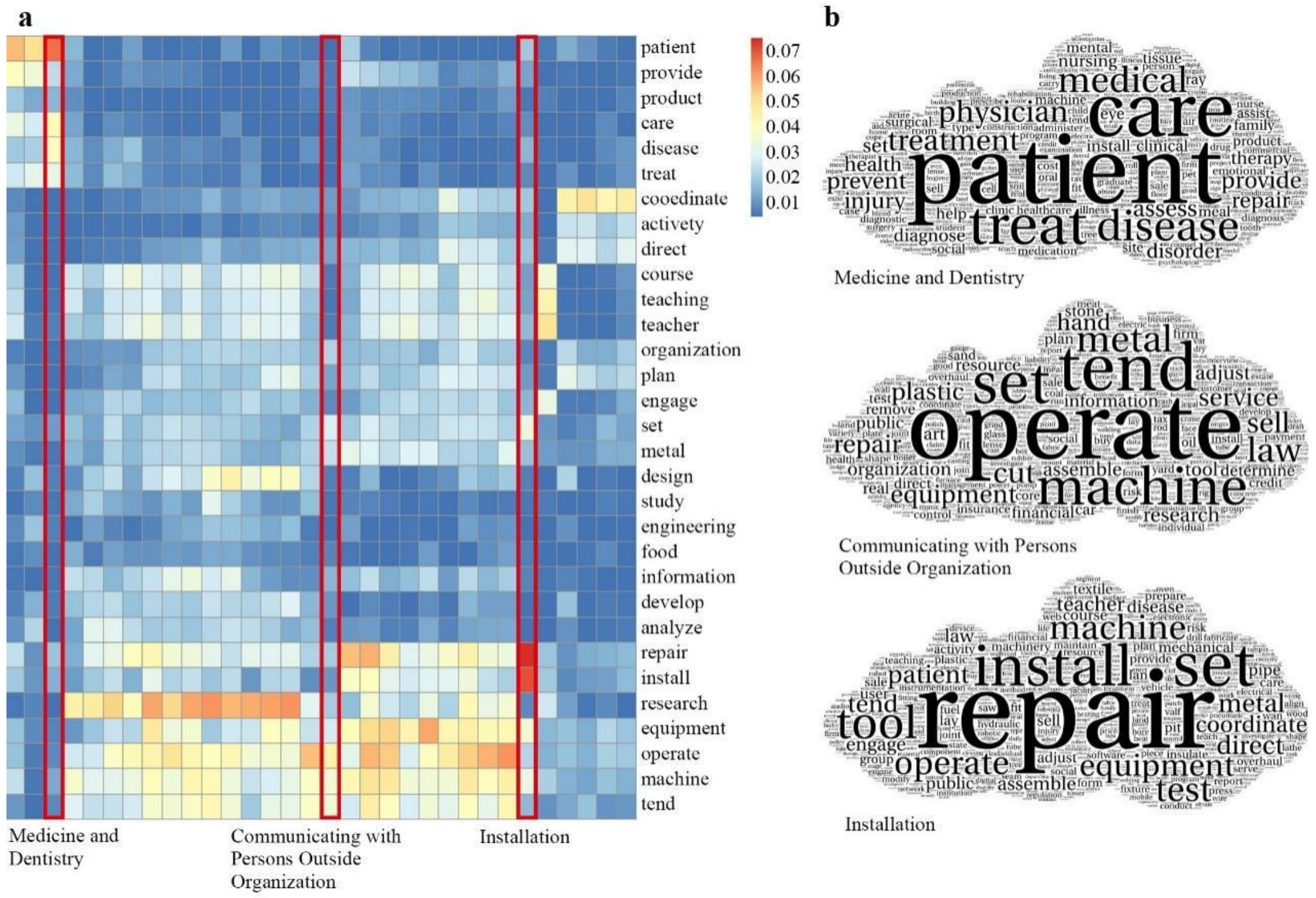

**Fig. S1 | Mutual information between O*NET tasks and skills. a**) The heat map is drawn based on the mutual information of the task-skill pairs, showing the heterogenous structure of the task-skill relationships. **b**) Task clouds of three skills: *Medicine and Dentistry*, *Communicating with Persons Outside Organization* and *Installation*. The word size is proportional to the value of the mutual information criterion.

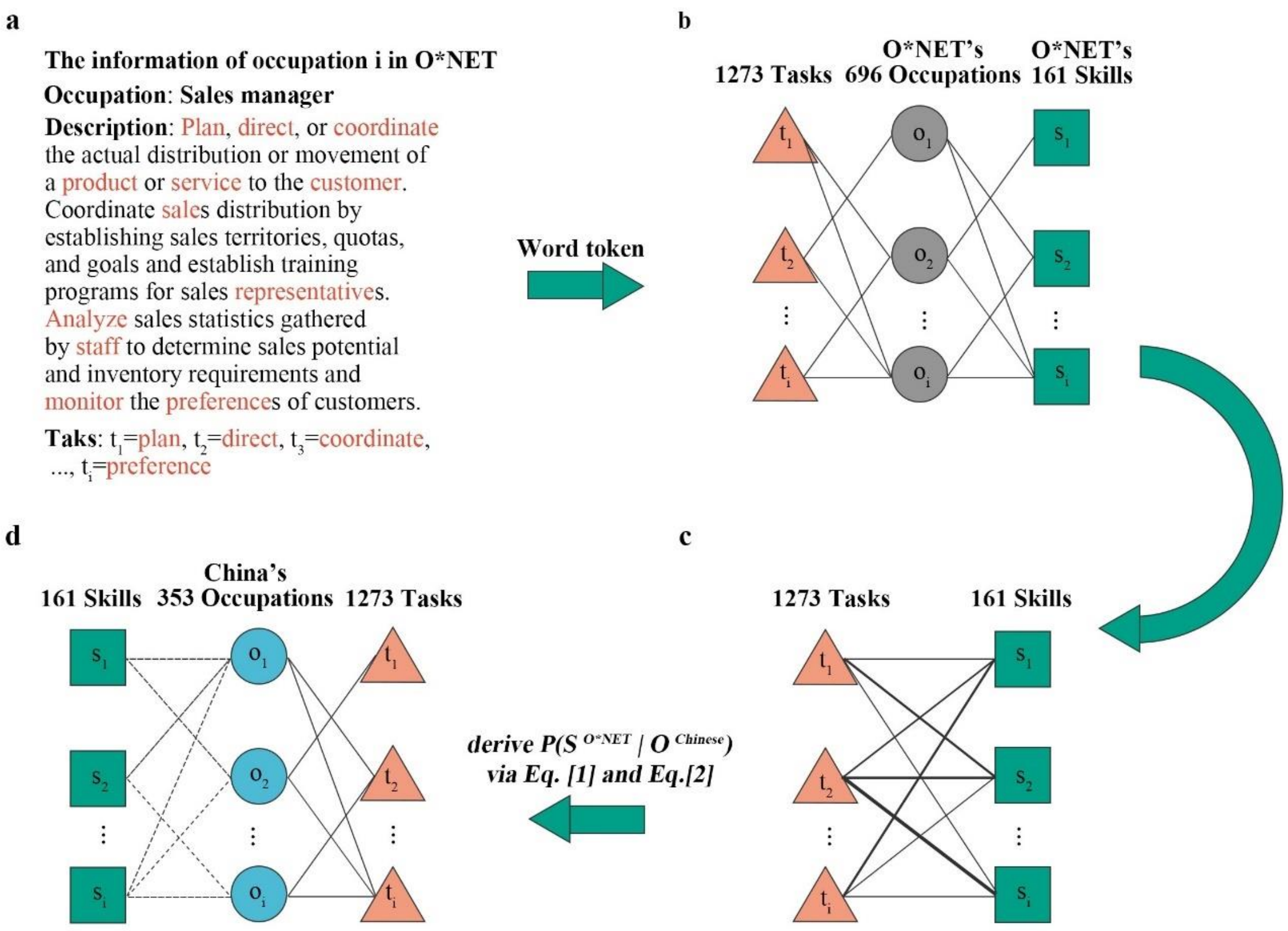

**Fig. S2 | The workflow of building China's first workforce skill taxonomy. a**) Sales manager's job descriptions in O*NET. We derive tasks by extracting word tokens from the job descriptions that are highlighted in red. All tasks can be found in Table. S1. **b**) A tripartite network links O*NET 1273 tasks, 161 skills and 696 occupations. **c**) The relationships between tasks and O*NET skills, represented by the conditional probability of whether or not a particular task $t_i$ would exist given a set of skills $s^{O*NET}$. **d**) Deriving the relationships between O*NET skills and Chinese occupations by using Naïve Bayesian inference.

## 4.4 City Skill Profile

Based on the workforce skill taxonomy obtained in this study, we compute the Chinese cities' skill profile, which is comprised of two parts: the numbers of skills effectively used and the socio-cognitive scores.

### The skills effectively used by Chinese cities

The determination of the skills effectively used by a city involves three steps. First, the skill taxonomy addresses whether or not a particular skill out of 161 skills is effectively used by any of the 353 occupations. We construct a matrix with 353 rows (occupations) and 161 columns (skills). The cell value of the matrix, 1 or 0, indicates whether the corresponding skill is effectively used by the corresponding occupation. Second, we address $CS(c, s)$ which represents the numbers of workers in city $c$ that use skill $s$ using Eq. [6], where $census(c, o)$ depicts the numbers of workers in city $c$ with occupation $o$ and $e(o, s)$ indicates whether or not skill $s$ is effectively used by occupation $o$. Finally, we can find out which skills are effectively used in the city by calculating the revealed comparative advantage ($RCA$) of each skill in 102 cities according to the Eq. [7]. If $RCA(c,s)$ is greater than 1, it means the skill $s$ is used effectively by city $c$. The higher the value of $RCA(c,s)$, the more the city $c$ has a comparative advantage in the skill $s$. Accordingly, the number of skills effectively used by city $c$ is $Skills_c$ (see Eq. [8])

$$CS(c,s) = \sum_{o \in O} census(c,o) * e(o,s) \quad [6]$$

$$RCA(c,s) = \frac{CS(c,s)/\sum_{s \in S} CS(c,s)}{\sum_{c \in C} cs(c,s)/\sum_{c \in C, s \in S} CS(c,s)} \quad [7]$$

$$Skills_c = \sum_{s \in S^{O*NET}} \begin{cases} 1, & RCA(c,s) \geq 1 \\ 0, & RCA(c,s) < 1 \end{cases} \quad [8]$$

## The socio-cognitive scores

The skills are divided into two clusters through community detection (see Fig. 4). There are 97 socio-cognitive skills and 64 sensory-physical skills among the 161 skills in total. The socio-cognitive level of an occupation can be defined as the percentage of socio-cognitive skills to the total number of cognitive skills, that is 97, see Eq. [9]. Taking the occupation entrepreneur for example, the number of the socio-cognitive skills is 66, so the socio-cognitive level is 0.68. Second, we determine whether or not an occupation is a socio-cognitive occupation by introducing a socio-cognitive level threshold. In this study, we take 0.6 as the threshold, all the occupations whose socio-cognitive level is higher than that are regarded as the socio-cognitive occupations while the others are non-socio-cognitive occupations. We also tried 0.7 and 0.8 as thresholds to test the robustness of the model, respectively (See Table S3). The findings hold in models using different thresholds. Finally, we determine a city's socio-cognitive scores $cognitive_c$ by the proportion of jobs that are socio-cognitive occupations to the total employment in the city (see Eq. [10]).

$$cognitive_o = \frac{\sum_{s \in Cognitive} e(o,s)}{97} \quad [9]$$

$$cognitive_c = \frac{\sum_{j \in cognitive_{o > threshold}} census(c,o)}{\sum_{o \in O} census(c,o)} \quad [10]$$

## Additional Reference

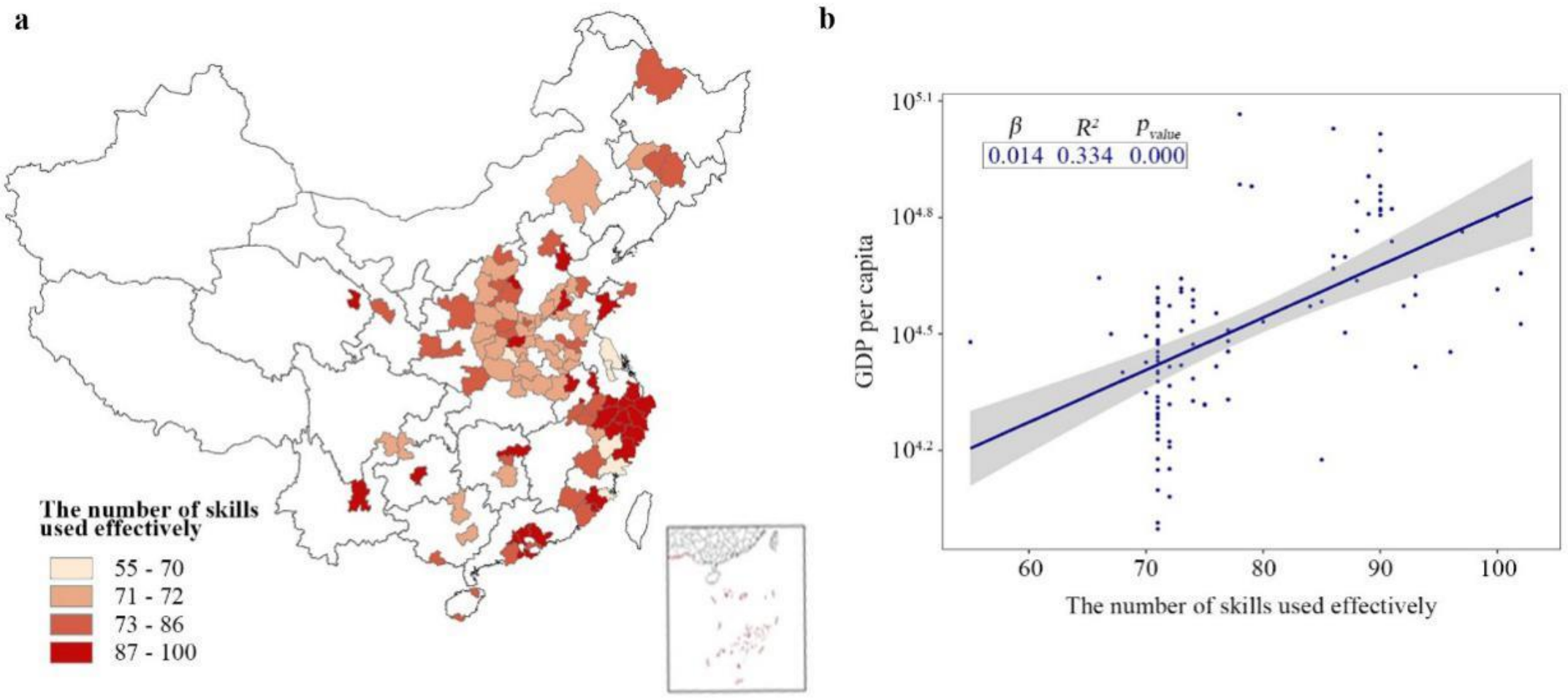

**Fig. S3 | The number of skills used effectively of 102 cities.** a) Cities are divided into quartiles. The colors address the numbers of skills effectively used by cities. b) The GDP per capita is related with the number of skills used effectively.

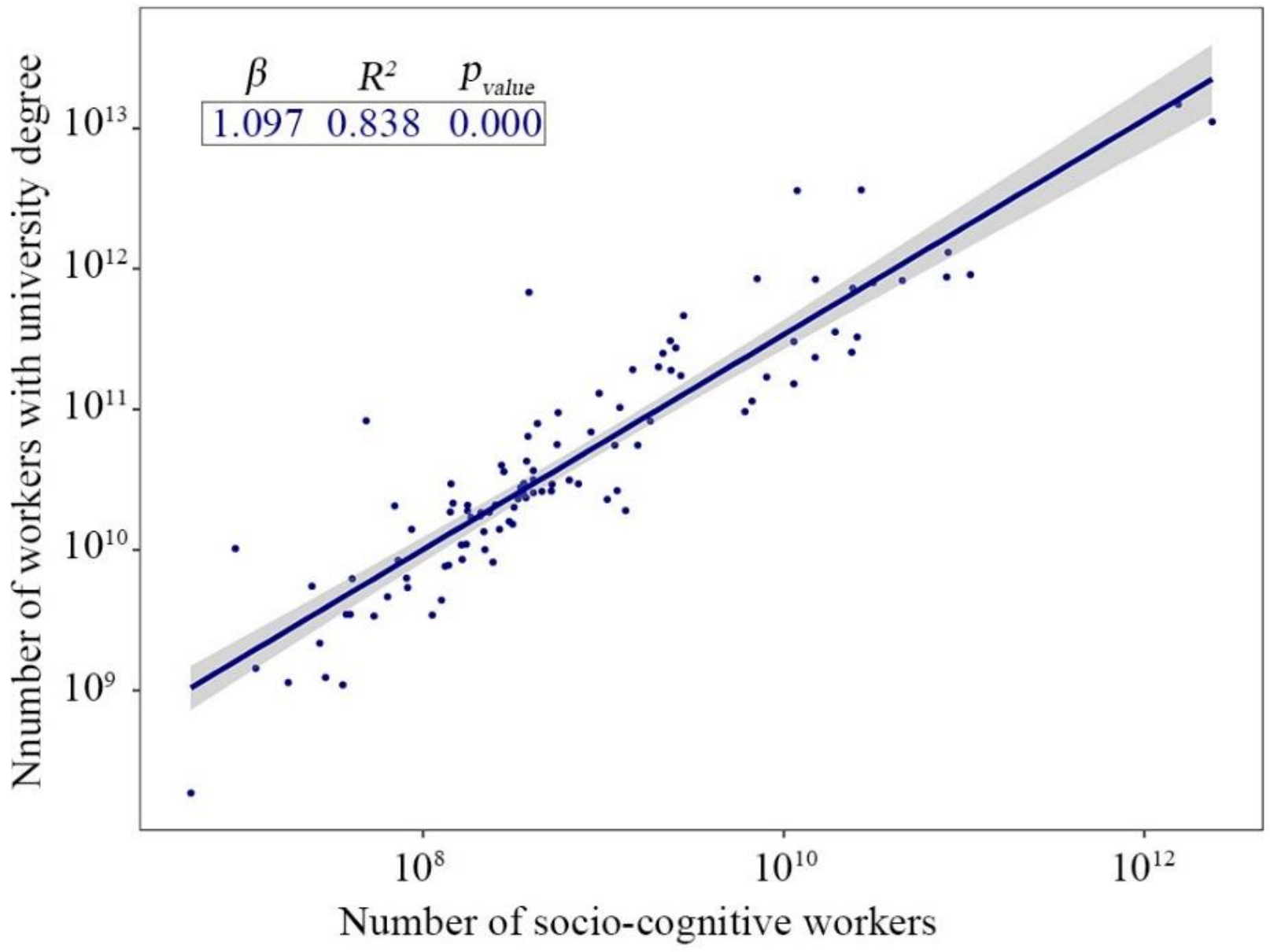

**Fig. S4 | The scatter diagram shows the relationship between the number of workers with a university degree and the number of socio-cognitive workers.**

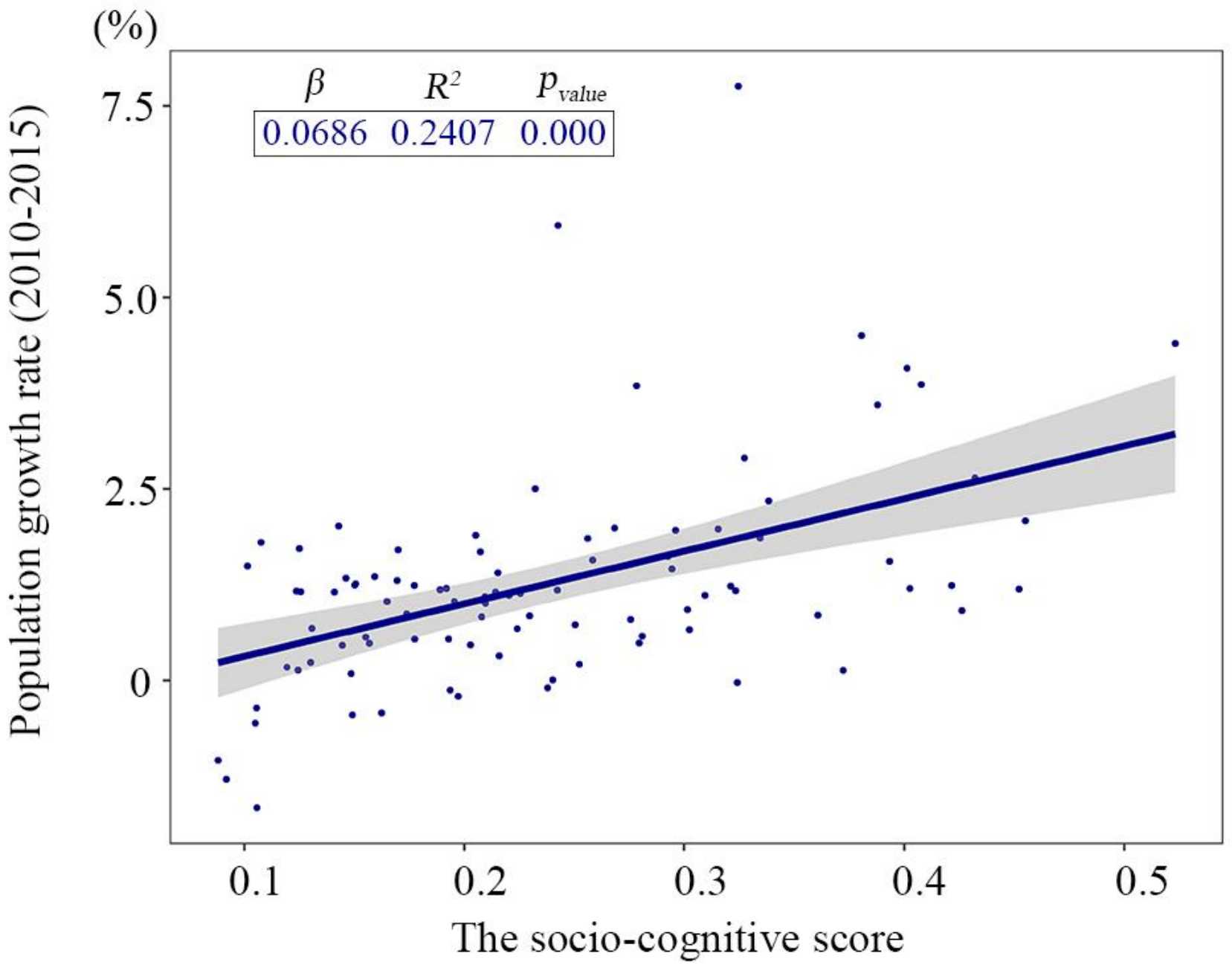

**Fig. S5| The scatter diagram shows a strong relationship between city's cognitive scores and the net population growth from 2010 to 2015.**